\documentclass[aps,pre,twocolumn,groupedaddress,amsmath,amssymb,superscriptaddress,reprint,showpacs,showkeys]{revtex4-1}

\usepackage{graphicx}
\usepackage{dcolumn}
\usepackage{color}
\usepackage{epsfig}
\usepackage{adjustbox}
\usepackage{subfigure}

\usepackage{amsmath}
\usepackage{multirow}
\usepackage{bm}
\usepackage{amsthm}
\usepackage{amsfonts}
\usepackage{float}
\usepackage{times}
\usepackage{amssymb}
\usepackage{relsize}
\usepackage{booktabs}
 
\usepackage[colorlinks=true,citecolor=blue,linkcolor=blue,urlcolor=blue]{hyperref}

\usepackage[sort&compress]{natbib}

\newcommand{\ket}[1]{\vert #1 \rangle}
\newcommand{\bra}[1]{\langle #1 \vert}

\begin{document}
	
\title{Light-matter quantum Otto engine in finite time}

\author{G. Alvarado Barrios}
\email[G. Alvarado Barrios]{\qquad phys.gabriel@gmail.com}
\affiliation{International Center of Quantum Artificial Intelligence for Science and Technology (QuArtist)\\  and Physics Department, Shanghai University, 200444 Shanghai, China}

\author{F. Albarr\'an-Arriagada}
\affiliation{International Center of Quantum Artificial Intelligence for Science and Technology (QuArtist)\\  and Physics Department, Shanghai University, 200444 Shanghai, China}

\author{F. J. Pe\~na}
\affiliation{Departamento de F\'isica, Universidad T\'ecnica Federico Santa Mar\'ia, Casilla 110V, Valpara\'iso 2340000, Chile}

\author{E. Solano}
\email[E. Solano]{\qquad enr.solano@gmail.com}
\affiliation{International Center of Quantum Artificial Intelligence for Science and Technology (QuArtist)\\  and Physics Department, Shanghai University, 200444 Shanghai, China}
\affiliation{Department of Physical Chemistry, University of the Basque Country UPV/EHU, Apartado 644, 48080 Bilbao, Spain}
\affiliation{IKERBASQUE, Basque Foundation for Science, Plaza Euskadi 5, 48009 Bilbao, Spain}
\affiliation{IQM, Nymphenburgerstr. 86, 80636 Munich, Germany}

\author{J. C. Retamal}
\email[J. C. Retamal]{\quad juan.retamal@usach.cl}
\affiliation{Departamento de F\'isica, Universidad de Santiago de Chile (USACH), Avenida Ecuador 3493, 9170124, Santiago, Chile}
\affiliation{Center for the Development of Nanoscience and Nanotechnology 9170124, Estaci\'on Central, Santiago, Chile}

\date{\today}

\begin{abstract}
We study a quantum Otto engine at finite time, where the working substance is composed of a two-level system interacting with a harmonic oscillator, described by the quantum Rabi model. We obtain the limit cycle and calculate the total work extracted, efficiency, and power of the engine by numerically solving the master equation describing the open system dynamics. We relate the total work extracted and the efficiency at maximum power with the quantum correlations embedded in the working substance, which we consider through entanglement of formation and quantum discord. Interestingly, we find that the engine can overcome the Curzon-Ahlborn efficiency when the working substance is in the ultrastrong coupling regime. This high-efficiency regime roughly coincides with the cases where the entanglement in the working substance experiences the greatest reduction in the hot isochoric stage. Our results highlight the efficiency performance of correlated working substances for quantum heat engines.
\end{abstract}

\maketitle

\section{Introduction}
A quantum heat engine (QHE) is a quantum device that generates power from the heat flow between a hot and a cold reservoir interacting with a quantum system. Therefore, the working mechanism of the engine is described by the laws of quantum mechanics. Quantum reciprocating heat engines operate under the quantum versions of well known thermodynamical cycles, such as Carnot or Otto cycles \cite{feldmann1996heat,feldmann2004characteristics,rezek2006irreversible, Henrich2007,PhysRevLett.93.140403,quan2007quantum,he2009performance}. The performance of these devices depend strongly on the choice of the working substance and the thermodynamic cycle which govern the dynamics. This has led to several cases being considered, for instance, noninteracting spin 1/2 particles, harmonic oscillators, atoms in harmonic traps, among others \cite{feldmann1996heat,feldmann2004characteristics,rezek2006irreversible, Henrich2007,PhysRevLett.93.140403,quan2007quantum,he2009performance,thomas2016performance,Watanabe2017quantum,He2002,feldmann2004characteristics,Geva1996,li2007quantum}. 

For a heat engine to achieve maximum performance, \textit{i.e.} maximum efficiency and energy extracted, it must employ infinitely long cycles since it demands thermalization of the working substance. As a natural consequence, the energy extracted per unit time, \textit{i.e.} the power, becomes vanishingly small. From a more realistic point of view, we would like to consider the operation of quantum heat engines for a finite cycle time, and find the optimal conditions where the energy extracted per unit time is maximum. For operation in finite time, four-stroke engines show a transient behavior from initialization to a periodic steady state, characterized by a sequence of states, referred to as the limit cycle \cite{feldmann1996heat,Insinga2016Thermo, feldmann2004characteristics}, which is usually achieved after a few cycles. The optimal performance of a cycle in  finite time is associated with the maximum power it can yield, which is obtained for an optimal cycle duration. The efficiency of the cycle in the optimal state is of particular relevance, it is simply termed efficiency at maximum power (EMP)  and, classically, it is limited by the Curzon-Ahlborn efficiency \cite{andresen1977thermodynamics,van2005thermodynamic}. This limit is also fulfilled for some quantum working substances  \cite{geva1992classical,esposito2009universality,wang2015efficiency}, but it is interesting to consider under what conditions a QHE can exceed this classical bound \cite{deffner2018}. The finite-time operation of the quantum Otto cycle has been studied for interacting and noninteracting working substances, such as, two-level systems \cite{geva1992quantum,feldmann2000performance}, harmonic oscillators \cite{geva1992classical,Kosloff2017Harmonic}, coupled harmonic oscillators \cite{wang2015efficiency} among others.  \par 
The possibility of extracting work from quantum correlations  \cite{Oppenheim2002Thermodynamical,Zurek2003QDandM} has highlighted the thermodynamic value of correlations. This  has motivated the study of the role of quantum correlations in the performance of QHEs whose working substance is composed by coupled quantum systems. For an appropriately strong coupling it is possible to have non-zero quantum correlations even when the global system is in thermal equilibrium with a reservoir at a given temperature. This problem has been studied in the context of QHEs in which the heat strokes act for a sufficiently long time so as to allow for a full thermalization \cite{Zhang2007,Zhang2008,Wang2009,dillenschneider2009,Altintas2014,Hardal2015,Alvarado2017Role}. From these works, it is clear that the effect of quantum correlations on the performance of quantum heat machines are model dependent. On the other hand, the role played by quantum correlations in finite time operation of QHEs is not yet completely understood. \par 

In this work, we consider a quantum Otto engine in finite time with a working substance composed by a two-level system (TLS) interacting with a harmonic oscillator, described by the quantum Rabi model (QRM), a paradigmatic light-matter interaction. Such working substance can be realized in superconducting  circuits \cite{niemczyk2010circuit,FornDiaz2010BSS,Bourassa2009USC,yoshihara2017DSC,forn2017ultrastrong,Yoshihara2017characteristic} where the interaction can be engineered to have access to different coupling regimes. We obtain the limit cycle and calculate the total work extracted, efficiency, and power of the engine by numerically solving the master equation describing the open system dynamics. Afterwards, we consider the relation between the light-matter quantum correlations and the total work extracted, efficiency and power, by considering entanglement of formation and quantum discord. We consider the reduction of quantum correlations during the isochoric stages and find that the engine can overcome the Curzon-Ahlborn efficiency when the working substance is in the ultrastrong coupling regime. In addition, we find that quantum discord, rather than entanglement acts as an indicator of the total work extracted by the engine.
 This may suggest that for this model, quantum correlations other than entanglement affect the performance of the engine.
 
\section{The model}
We consider a QHE where a TLS is coupled to a single quantized mode, described by the QRM dipolar coupling
\begin{equation}
\label{RabiHamiltonian}
H = \hbar \omega_{q} \sigma_{z} + \hbar \omega_{r} a^{\dagger}a + \hbar g \sigma_{x} (a + a^{\dagger}) .
\end{equation} 
Here, the operators $\sigma_{\alpha}$, $\alpha = \{x, z \}$, are the Pauli matrices corresponding to the TLS, and $a(a^{\dag})$ is the annihilation (creation) bosonic operator for the harmonic mode. The parameters, $\omega_{r}$, $\omega_{q}$ and $g$, describe the resonator frequency, qubit frequency, and qubit-resonator coupling strength, respectively. Throughout this work we will consider the resonance condition $\omega_{ r } = \omega_{q} = \omega$.

The dynamics of the QRM can be separated into three different regimes which are governed by the ratio $g/\omega$ \cite{Wolf2013,Rossatto2016}. When $g$ is much larger than any decoherence or dephasing rate in the system, and $g/\omega \lesssim 10^{-1}$ we can define the strong coupling (SC) regime. In this regime, the rotating wave approximation (RWA) can be performed, obtaining the Jaynes-Cummings model \cite{JC1963}, in this model the number of excitations is preserved, thus, the dynamics involve only states with the same number of excitations as in the initial state. For $0.1 < g/\omega \leq 1$ is considered the ultrastrong coupling regime (USC)
 \cite{niemczyk2010circuit,FornDiaz2010BSS,Bourassa2009USC}. In  the USC regime, the RWA cannot be performed, then, the dynamics of the QRM no longer preserve the number of excitations. Finally, when the coupling strength is larger than the frequency of the system, \textit{i.e.} $g \gg \omega$ we obtain the deep-strong coupling regime (DSC) \cite{yoshihara2017DSC,Casanova2010DSC}. In this regime again the TLS degrees of freedom decouple from the harmonic oscillator degrees of freedom \cite{DeLiberato2014PRL}. 

One  characteristic of the spectrum of the QRM is that the energy levels start to degenerate when the coupling parameter is increased from the SC regime to the DSC regime. Also, in the USC the eigenstates of the model correspond to highly correlated states of the TLS and resonator mode, ideal to study the role of quantum correlation of the working substance in the performance of a QHE. \par

\section{Quantum Otto cycle}  \label{QOE}
In what follows we study a QHE operating under a quantum Otto cycle in finite time. This cycle is composed of two adiabatic processes and two isochoric processes. In the latter, the working substance interacts with a cold(hot) reservoir at temperature $T_{c}$($T_{h}$). We will consider that the coupling strength $g$ is kept constant throughout the cycle. \par 

For the adiabatic processes, we consider a change in the resonator frequency between the values $\{\omega_{h}, \omega_{c}\}$ with $\omega_{h}>\omega_{c}$. The timescale, $\tau_{ad}$ of this process must be sufficiently large as to satisfy the adiabatic theorem \cite{Born1928adiabatictheorem}. This, in general, is satisfied if the duration is much larger than the dynamical timescale, that is, $\tau_{ad} \gg \hbar / \varepsilon$ \cite{Born1928adiabatictheorem}, where $\varepsilon$ is a characteristic transition energy of the system, in this case, it can be considered to be of the order of GHz. This choice is justified as long as the smallest relevant transition frequency remains in the range of GHz. \par

For the isochoric processes, we consider that the system undergoes a Markovian evolution, described by \cite{Blais2011}

\begin{equation}
\dot{\rho}(t) = -i[H,\rho(t)] + \sum_{m}\mathcal{L}_{m}(\rho(t)).
\end{equation}

Where

\begin{figure}[t!]
	\includegraphics[width=0.9\linewidth]{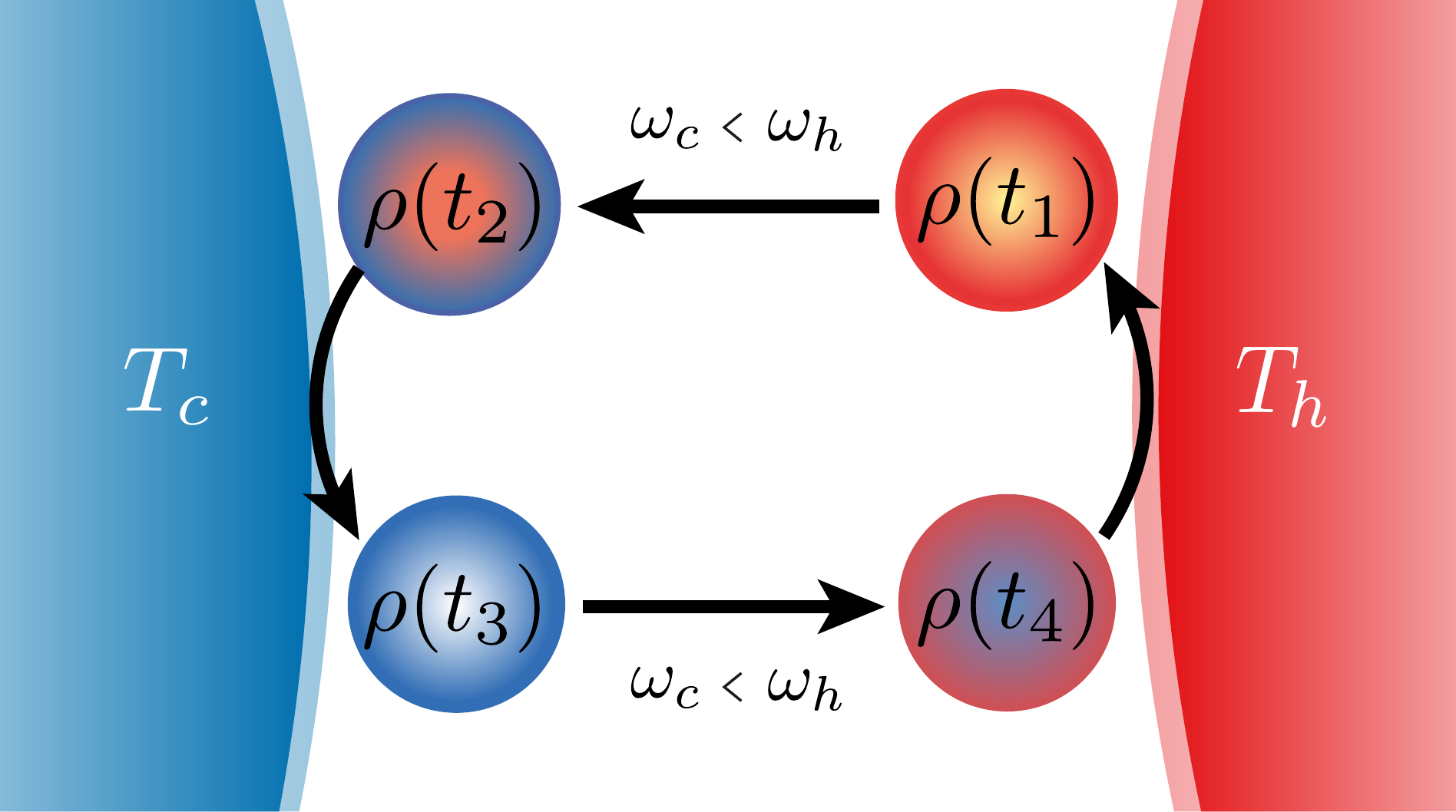}
	\caption{Diagram of the considered quantum Otto cycle.}
	\label{fig:ottodiagram}
\end{figure}

\begin{eqnarray}
\mathcal{L}_{m}(\rho(t)) = && \sum_{j, k>j} \Gamma_{m}^{jk} \bar{n}(\Delta_{kj},T) D(\ket{k}\bra{j})\rho(t) \nonumber \\ 
&& + \sum_{j,k>j} \Gamma_{m}^{jk} (\bar{n}(\Delta_{kj},T) + 1) D(\ket{j}\bra{k}) \rho(t). \nonumber \\
\end{eqnarray}

Where $\{\ket{j} \}_{j=0,1,2..}$ are the eigenvectors of the Hamiltonian $H$ of Eq.(\ref{RabiHamiltonian}), obeying $H\ket{j}= \epsilon_{j}\ket{j}$, and ${D(O)\rho = 1/2(O \rho O^{\dagger} - O^{\dagger}O \rho - \rho O^{\dagger}O)}$. The decay rates $\Gamma_{m}^{jk}$ are taken as \cite{Ridolfo2012}

\begin{equation}
\Gamma_{m}^{jk}  = \kappa_{m} \frac{\Delta_{kj}}{\omega_{0}}|C_{kj}^{m}|^{2},
\end{equation}

 where $m = 1$ refers to the qubit $m = 2$ to the resonator, $\Delta_{kj} = \epsilon_{k} - \epsilon_{j}$, ${\bar{n}(\Delta_{kj},T) = 1/\big(\exp(\frac{\hbar\Delta_{kj}}{k_{B}T})-1\big)}$ and ${C_{kj}^{m} = -i\bra{k} C^{m} \ket{j}}$, with $C^{m=1} = \sigma_{x}$ and $C^{m=2} = a + a^{\dagger}$. Here we have only considered energy relaxation in the master equation.\par

 For both the qubit and resonator dissipation channels, we consider the decay rates as $\kappa_{m} \sim 10^{-3}\omega$, $\{m = 1, 2 \}$ \cite{Ridolfo2012}. Notice that the relaxation timescale of the system is of the order of MHz. On the other hand, the dynamical timescale of the system, $\hbar/E$, is in the order of nanoseconds. Therefore, the total duration of the cycle is dominated by the relaxation timescale.\par  
 
 The thermodynamic cycle that we consider is shown in Fig.\ref{fig:ottodiagram} and proceeds as follows:\par 

Stage 1: Quantum isochoric process (hot bath stage)\cite{quan2007quantum}. The system, with frequency $\omega = \omega_{h}$ and Hamiltonian $H_h$, prepared in some initial state $\rho_{i}$, is brought into contact with a hot thermal reservoir at temperature $T_h$ for a time duration $\tau_{1}$. At the end of this process the state becomes $\rho(\tau_{1})$. During this process only the populations change while the energy level structure remains invariant.\par

Stage 2: Quantum adiabatic (expansion) process \cite{PhysRevLett.93.140403, quan2007quantum}. The system is isolated from the hot reservoir, and the resonator frequency is changed from $\omega_{h}$ to $\omega_{c}$, with $\omega_{h}>\omega_{c}$, to satisfy the quantum adiabatic theorem \cite{Born1928adiabatictheorem,messiah1958quantum}, the duration of this process, $\tau_{2}$, is chosen as ${\tau_{2} \gg \hbar/(\epsilon_{2}^{h} - \epsilon_{1}^{h})}$, where $\epsilon_{i}^{h(c)}$ is the i-th eigenenergy of Hamiltonian $H_{h(c)}$. 
During this process only the energy level structure changes. We denote the Hamiltonian at the end of this process by $H_c$, and the state of the system is $\rho(\tau_{2})$.\par

Stage 3: Quantum isochoric process (cold bath stage). The working substance, with $\omega = \omega_{c}$ and Hamiltonian $H_c$, is brought into contact with a cold thermal reservoir at temperature $T_c$ for a time duration $\tau_{3}$. At the end of this process, the state of the system is $\rho(\tau_{3})$. Since the resonator frequency has changed to $\omega_{c}$ due to the adiabatic process, in this stage the ratio $g/\omega_{c}$ for a given value of $g$ is different than in stage 1 as a consequence of the adiabatic process.\par 

Stage 4: Quantum adiabatic (compression) process. The system is isolated from the cold reservoir, and the resonator frequency is changed from $\omega_{c}$ to $\omega_{h}$. During this process the populations remain constant while the energy level structure returns to its configuration in Stage 1. The duration of this process, $\tau_{4}$, is chosen as ${\tau_{4} \gg \hbar/(\epsilon_{2}^{c} - \epsilon_{1}^{c})}$. At the end of this process the state of the system is given by $\rho(\tau_{4})$.\par 

The total cycle time, $\tau$, is the sum of the time duration of each stage $\tau = \sum_{i} \tau_{i}$. We will consider equal time duration for the isochoric stages, $\tau_{1} = \tau_{3}$, then $\tau = 2\tau_{1} + \tau_{2} + \tau_{4}$.\par  

Since we are considering finite duration for the thermodynamic cycle, the initial state of the system has an effect on the performance of the QHE. We will consider that the system starts in a thermal state at temperature $T_{c}$, and furthermore, we study the QHE under several consecutive iterations of the cycle. Since the isochoric stages involve a Markovian evolution, we can expect that after several iterations, there should be a limit cycle which is independent of the initial state of the system. \par

To describe a thermodynamical cycle in finite time requires the formulation of thermodynamic laws in a dynamical context \cite{kosloff2013quantum,Gelbwaser2015}. Let us consider a dynamical system with a discrete time-dependent Hamiltonian $H(t)$. The average energy of the system, $U(t)$, for a state $\rho(t)$ is given by 

\begin{equation}
U(t) = \langle H(t) \rangle = Tr\{\rho(t) H(t)\}.
\end{equation}

The rate of change of the average energy of the system is given by

\begin{equation}
\frac{dU}{dt} = \frac{dW}{dt} + \frac{dQ}{dt}.
\end{equation} 

This expression is a dynamical form of the first law of thermodynamics. The quantities $Q$ and $W$ are associated to heat and work, respectively, and can be written as

\begin{equation}
Q(t) = \int_{0}^{t} dt' Tr\{H(t') \dot{\rho}(t')\},
\label{heatdef}
\end{equation}

and

\begin{equation}
{W(t) = \int_{0}^{t} dt' Tr\{\dot{H}(t') \rho(t')\}}.
\label{workdef}
\end{equation}

Furthermore, we can write the power, $P$, as

\begin{equation}
P(t) = \frac{dW}{dt} = Tr\{\dot{H}(t) \rho(t)\}.
\end{equation}

Notice that for an isochoric process we have $H(t) = H$, which means $W(t)=0$. On the other hand, in an adiabatic process the system undergoes a unitary evolution with a time-dependent Hamiltonian, which naturally means $Q(t) = 0$.\par 

The quantum Otto cycle consists of four stages, in which the energy change of the system is accounted purely by work or purely by heat. Since the cycle is completed by returning to the initial configuration of the system, we can write the total work extracted as $W(t) = Q_{1}(t) + Q_{3}(t)$, where the subindex indicates the stage of the cycle as explained above.\par 

The efficiency of the engine is defined as the ratio of the total work extracted and the heat enters the system

\begin{equation}
\eta(t) = \frac{W(t)}{Q(t)}.
\end{equation}

The Carnot efficiency $\eta_{C} = 1 - \frac{T_{c}}{T_{h}}$ represents a universal upper-bound for all heat engines. However, the efficiency bound for a working cycle at maximum power is usually of greater practical importance than the Carnot efficiency. The efficiency at maximum power obeys the Curzon-Ahlborn (CA) bound \cite{andresen1977thermodynamics, van2005thermodynamic, deffner2018}, given by

\begin{equation}
\eta_{\text{CA}} = 1 - \sqrt{ \frac{T_{c}}{T_{h}}}.
\end{equation} 

\section{Limit Cycle} 

\begin{figure}[t!]
	\includegraphics[width= 0.99 \linewidth]{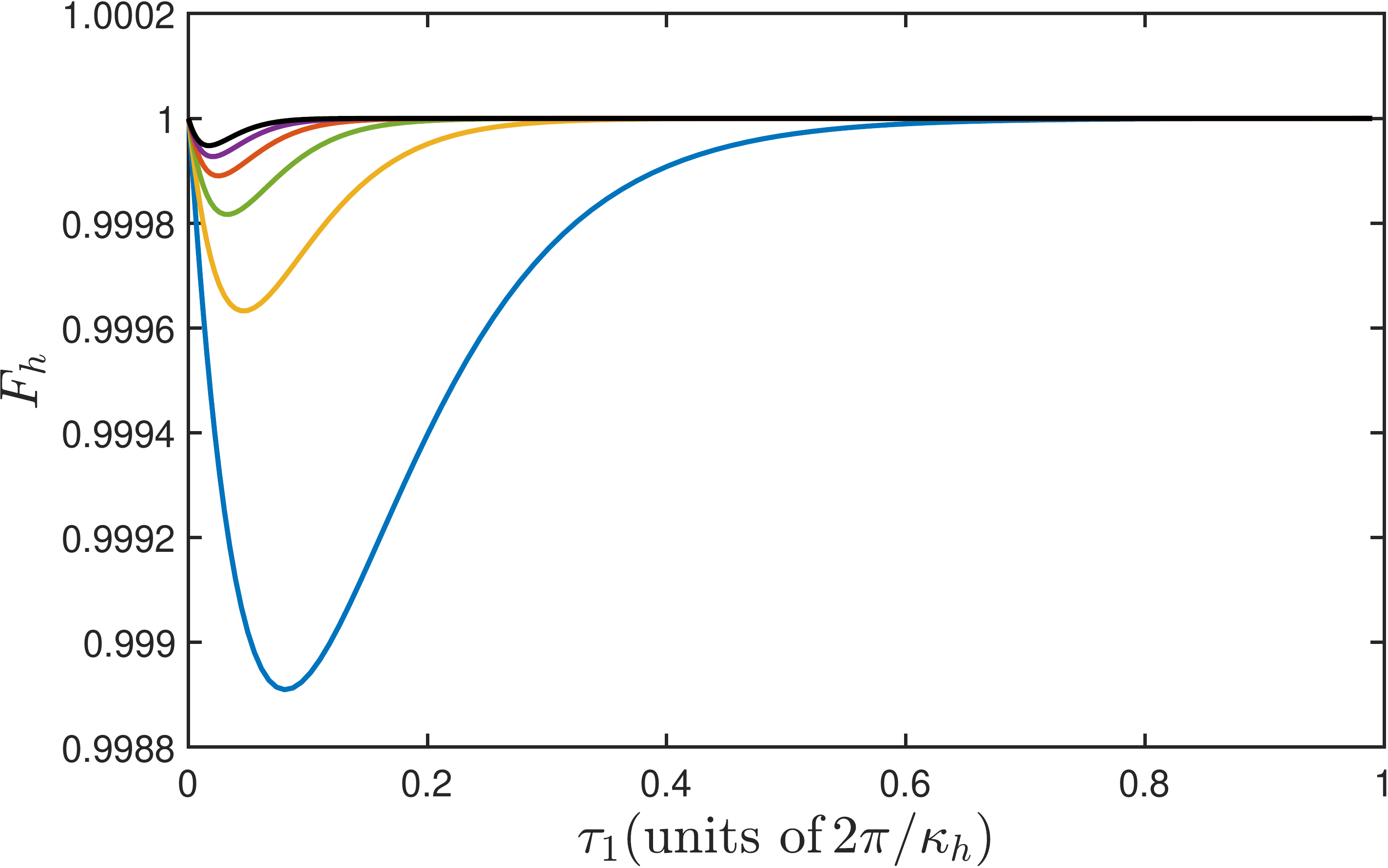}  
	\caption{(Fidelity between $\rho^{(N)}(\tau_{1})$ and $\rho^{(N-1)}(\tau_{1})$ in terms of $\tau_{1}$ for different iterations, where  $N = 2$ (blue), $N=3$ (yellow), $N = 4$  (green), $N = 5$ (red) and $N=6$ (black). We have chosen ${g/\omega_{c} = 0.5}$, $T_{c}=20$ mK and $T_{h}=8T_{c}$.}
	\label{fig:CycIter}
\end{figure}

Let us first consider the behavior of the engine over several iterations of the cycle. We will denote by $\rho^{(N)}(\tau_{1})$ the state of the system at the end of the hot isochoric stage for the Nth iteration of the cycle. We consider that each iteration of the cycle is identical. In Fig.~\ref{fig:CycIter} we plot the fidelity, $F_{h} = F(\rho^{(N-1)}(\tau_{1}),\rho^{(N)}(\tau_{1}))$, of state $\rho(\tau_{1})$ between consecutive iterations. As can be seen, when $\tau_{1}\rightarrow 0$ the state of the system experiences almost no change during the cycle, and therefore we have $F_{h} \rightarrow 1$. For  $0 < \tau_{1} \ll 1$, the evolution through one full cycle changes the state of the system leading to a decrease in the fidelity between consecutive cycles. As we consider increasing values of $\tau_{1}$, the interaction with the hot reservoir drives $\rho(\tau_{1})$ closer to the stationary thermal state, and the fidelity increases with $\tau_{1}$. For times comparable with the relaxation time, $F_{h} \rightarrow 1$ which indicates that the engine operates in the limit cycle from the first iteration. As expected, the minimum cycle time required to achieve maximal fidelity diminishes as we increase the number of iterations.

 We want to characterize the performance of the engine regardless of the state in which it is initialized, to this end, we consider the thermodynamic figures of merit in the limit cycle.\par 

\begin{figure}[t!]
	\includegraphics[width=0.99 \linewidth]{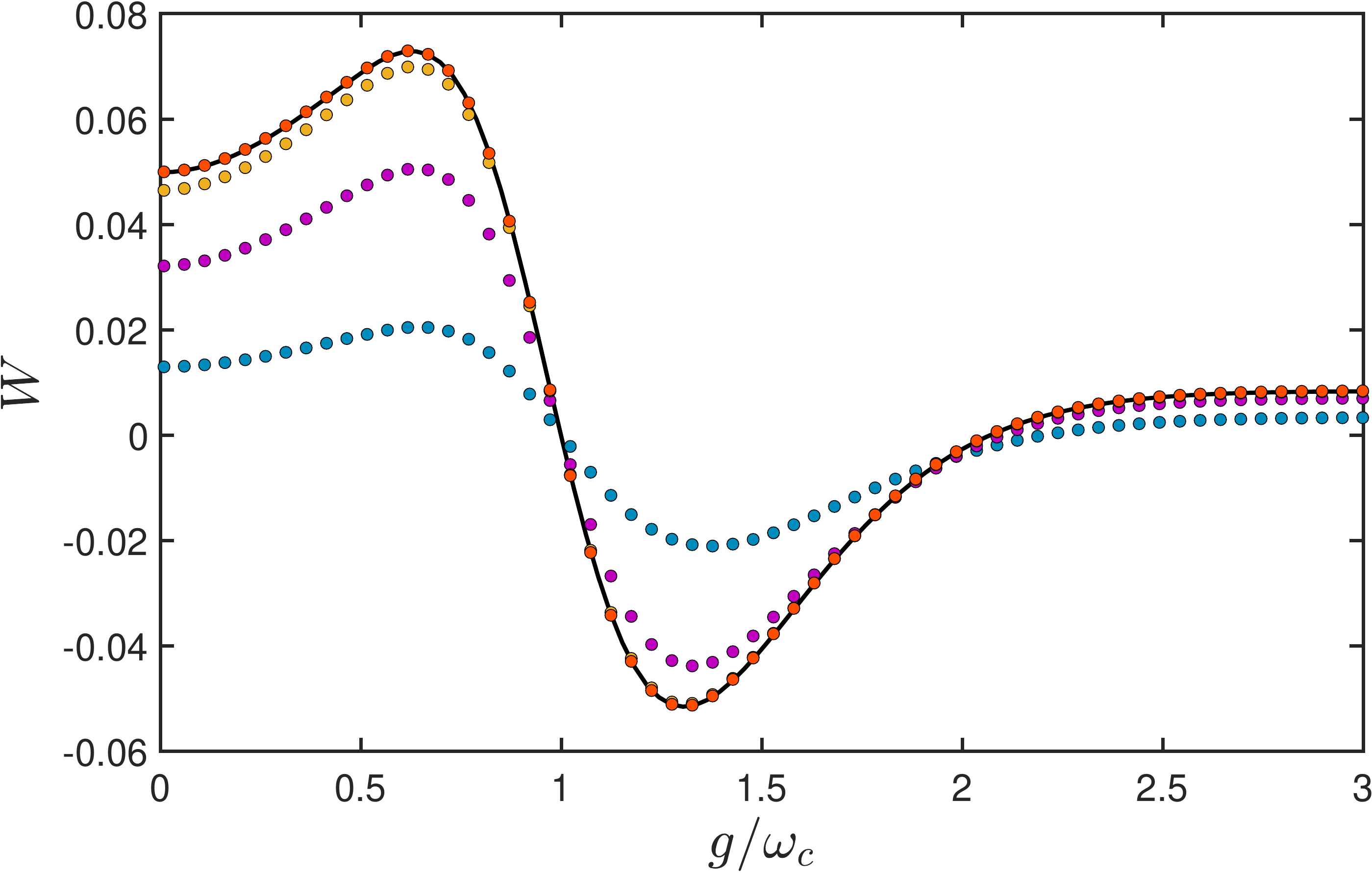}	
	\caption{Total Work extracted as a function of the coupling strength~$g/\omega_{c}$. The dots denote the total work extracted for finite cycle times for $\tau = 0.1$ (blue), $\tau = 0.3$ (purple), $\tau = 0.7$  (yellow) and $\tau = 1.0$ (red) in units of $2\pi/\kappa_{h}$. The black solid line is the total work extracted for a stationary cycle (thermalizing working substance). We have considered $T_{c} = 20$ mK and $T_{h} = 8T_{c}$.}
	\label{Fig:Comparison}
\end{figure}

Now, let us consider, for the limit cycle, the dependence of the total work extracted on the coupling strength $g/\omega_{c}$, for different cycle times. This is shown in Fig.~\ref{Fig:Comparison}, where the cycle times differ mainly in the time assigned to the isochoric stages, $\tau_{1}$ and $\tau_{3}$ ($\tau_{1} = \tau_{3}$). As can be seen in the figure, the total work extracted increases with the cycle time until the system closely approaches thermalization with the heat reservoirs in the isochoric stages. This is because when the duration of the isochoric stages are large compared to the relaxation times, the thermodynamic quantities reach their corresponding values for a stationary Otto cycle \cite{Alvarado2017Role}. \par 

\begin{figure}[t!]
	\includegraphics[width=0.9\linewidth]{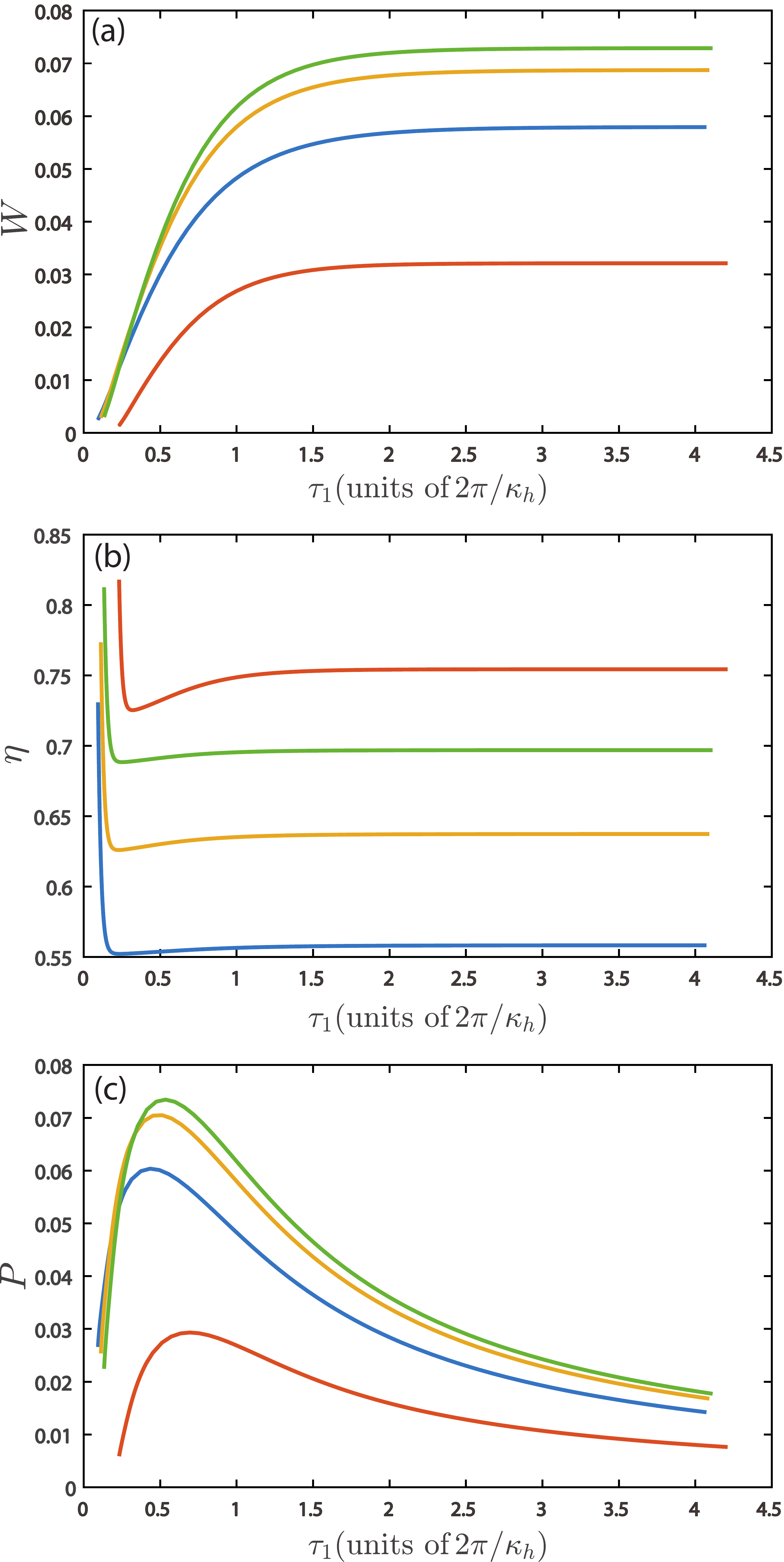}	
	\caption{(a) Total work extracted, (b) efficiency and (c) power as a function of the total cycle time $\tau$ for $g/\omega_{c} = 0.3$ (blue), $g/\omega_{c} = 0.5$ (yellow), $g/\omega_{c} = 0.63$  (green) and $g/\omega_{c} = 0.9$ (red).}
	\label{Fig:WEP}
\end{figure}

 In Fig.~\ref{Fig:WEP}, we show the total work extracted, efficiency and power as a function of the total cycle time $\tau$ for different values of $g/\omega_{c}$. Since the adiabatic processes have finite duration to satisfy the adiabatic approximation, the total cycle time cannot be less than $\tau = \tau_{2} + \tau_{4}$. Where $\tau_{2}$ and $\tau_{4}$ depend on $g/\omega_{c}$. This can be seen in Fig.~\ref{Fig:WEP}, where we see that for each quantity the smallest values of $\tau$ increase with $g/\omega_{c}$. We see that when the adiabatic times do not change significantly with $g/\omega_{c}$ and for very small cycle times, the coupling strength has little effect on the total work extracted. On the other hand, we can see that for large cycle times the total work extracted approaches its stationary value, matching the amount in Fig.~\ref{Fig:Comparison} for corresponding values of $g/\omega_{c}$. In Fig.~\ref{Fig:WEP}(b) we see that the efficiency is greater for increasing values of the coupling strength. In Fig.~\ref{Fig:WEP}(c) we can see that there is an optimal cycle time that maximizes the energy extracted per unit time. This optimal performance depends on the cycle duration and the strength of the coupling parameter. Thus, the more relevant quantities that characterize the quantum Otto cycle in finite time are the power and the EMP. In the following sections we will analyze the relationship between these quantities and the quantum correlations between the TLS and the resonator, embedded in the working substance. \par

\begin{figure}[t!]
	\includegraphics[width=0.9 \linewidth]{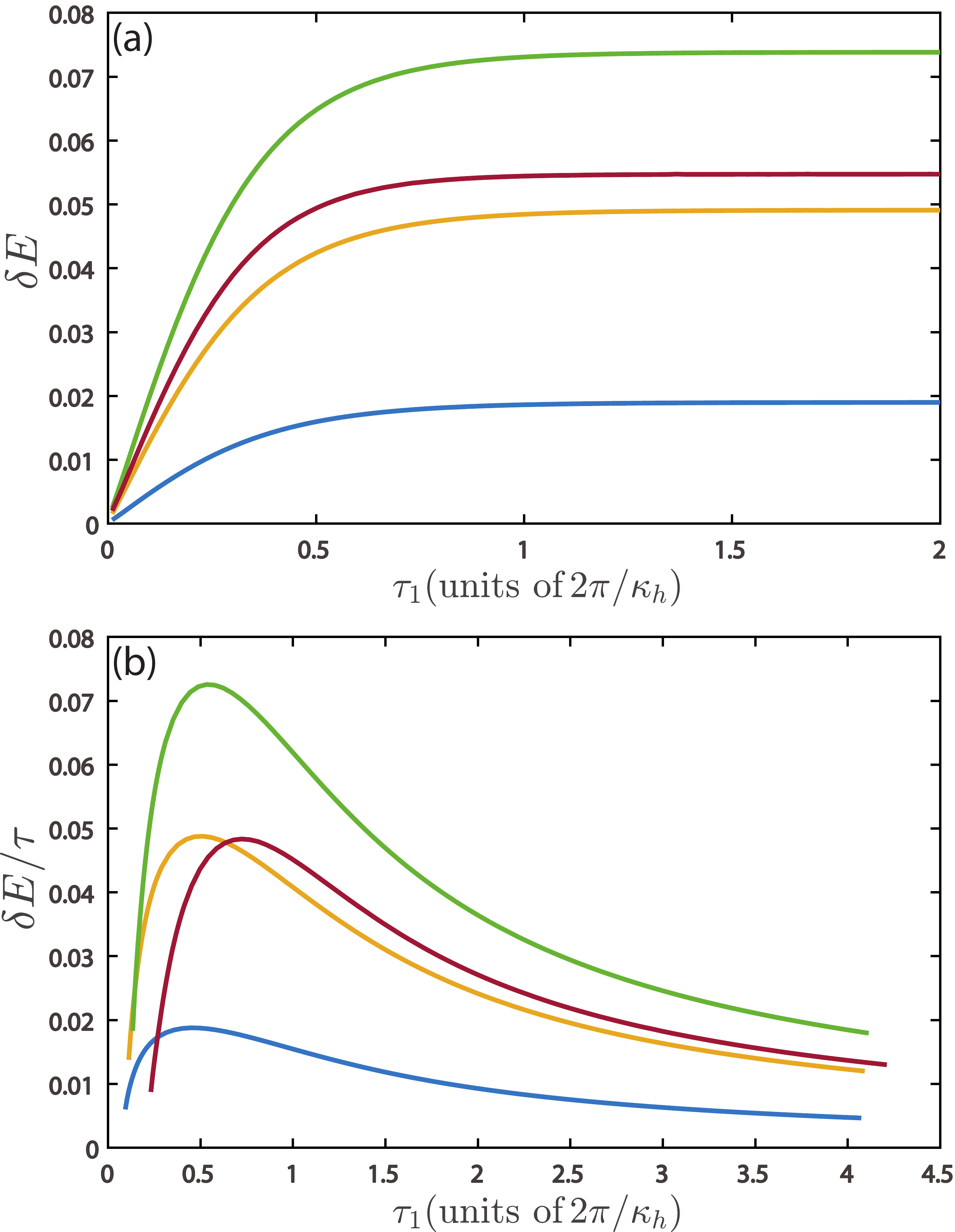}
	\caption{Difference of entanglement $\delta E = E(\rho(\tau_{4})) - E(\rho(\tau_{1}))$ as a function of the interaction time with the hot thermal reservoir, $\tau_{1}$, for the limit cycle, for $g/\omega_{c} = 0.3$ (blue), $g/\omega_{c} = 0.5$ (yellow), $g/\omega_{c} = 0.63$  (green) and $g/\omega_{c} = 0.9$ (red). We have chosen ${T_{h} = 160 \textrm{mK}}$.}
	\label{Fig:Ent}
\end{figure}

\section{Quantum Correlations for finite-time operation}
An interesting issue is to consider the effect of quantum correlations between the components of the working substance in the performance of the QHE. This effect has been studied previously in this context for an Otto cycle in which the system thermalizes during the isochoric stages (infinite time QHE)~\cite{Hardal2015,Alvarado2017Role}. These studies have shown that the change in quantum correlations during the hot isochoric stage are indicative of the behavior of the total work extracted. However, the role that these quantum correlations may play in the performance of finite-time Otto cycles has not been addressed.\par
We will study the quantum correlations between the two-level system and the resonator. We will quantify two kinds of quantum correlations, namely, the entanglement of formation (EoF) \cite{Bennet1996EOF} and quantum discord (QD) \cite{Zurek2001QuantumDiscord}.\par
For a general mixed state $\rho_{AB}$ of a bipartite system $A$ and $B$, we can write a specific pure state decomposition as $\rho_{AB} = \sum_{i}p_{i}|\Phi_{i}\rangle\langle\Phi_{i}|$. We can define the entanglement of this pure state decomposition as

\begin{equation}
E_{p_i,\Phi_i} = \sum_{i} p_{i} E(|\Phi_{i}\rangle), 
\label{ecu}
\end{equation}

where $E(|\Phi\rangle) = S(\rho_{A(B)})$ is the entanglement of the pure state $|\Phi_{i}\rangle$, $S(\mathcal{A}) = -\textrm{tr}(\mathcal{A}\ln \mathcal{A})$ is the von Neumann entropy, and $\rho_{A(B)}$ is the reduced density matrix of the subsystem $A(B)$. Now, the EoF of the density matrix $\rho$ is given by a minimization over all the possible pure state decompositions ${p_i,\Phi_i}$ of Eq. (\ref{ecu}), that is

\begin{figure}[t!]
	\includegraphics[width=0.9 \linewidth]{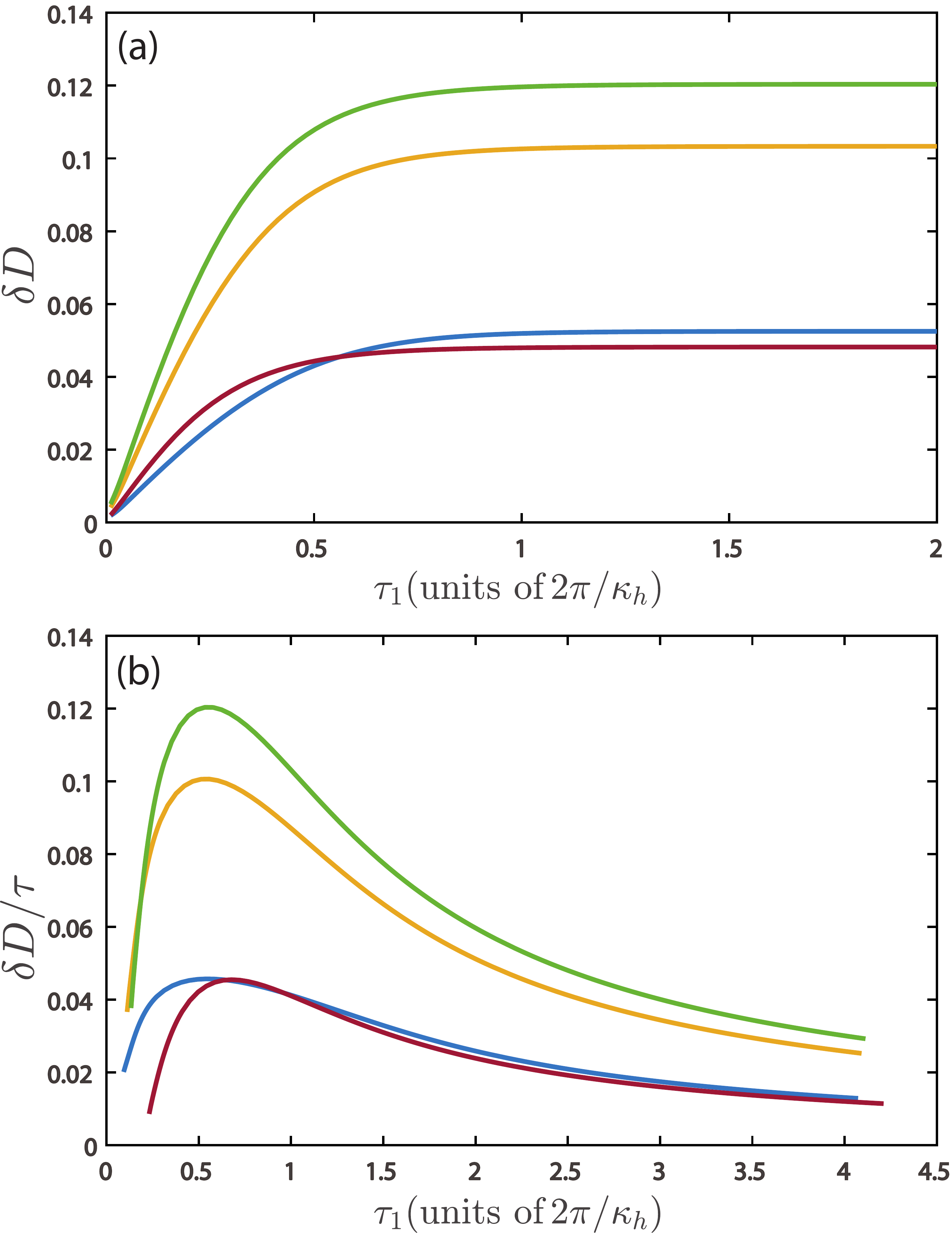}
	\caption{Difference of quantum discord $\delta D = D(\rho(\tau_{4})) - D(\rho(\tau_{1}))$ as a function of the interaction time with the hot thermal reservoir, $\tau_{1}$, for the limit cycle, for $g/\omega_{c} = 0.3$ (blue), $g/\omega_{c} = 0.5$ (yellow), $g/\omega_{c} = 0.63$  (green) and $g/\omega_{c} = 0.9$ (red). We have chosen ${T_{h} = 160}$ mK.}
	\label{Fig:QD}
\end{figure}

\begin{equation}
E(\rho) = \underset{p_{i},|\Phi_{i}\rangle}{\mathrm{min}} \sum_{i} p_{i} E(|\Phi_{i}\rangle), 
\end{equation}

it should be noted that EoF does not capture all forms of quantum correlations. On the other hand, quantum discord can quantify all the quantum correlations (other than entanglement).

The quantum discord in a bipartite $AB$ system can be calculated as follows \cite{Zurek2001QuantumDiscord}:

\begin{equation}
D_{A} = S(\rho_{A}) - S(\rho_{AB}) + \underset{\{\Pi_{j}^{A} \}}{\mathrm{min}} S(\rho_{B|\{\Pi_{j}^{A} \}}),
\end{equation}

where $\rho_{B|\{\Pi_{j}^{A} \}}$ is the state of the bipartite system after a projective measurement $\Pi_{j}^{A}$ has been performed on subsystem $A$. In our case, subsystem A will denote the  two-level system and subsystem B the resonator mode. Due to the size of the Hilbert space of the working substance we only consider projective measurements on the two-level system.\par 
To calculate either of the two measures it is necessary to solve a minimization problem, for which we employ the simulated annealing technique used in ref.~\cite{Allende2015simulated}. \par

Now, we analyze the relation between quantum correlations and the thermodynamic figures of merit of the cycle. We define the following quantities; we denote the difference of quantum correlations between the initial and final states of the hot isochoric stage as measured by entanglement by $\delta E = E(\rho(\tau_{4})) - E(\rho(\tau_{1}))$, and as measured by quantum discord by $\delta D =  D(\rho(\tau_{4})) - D(\rho(\tau_{1}))$ . In Fig.~\ref{Fig:Ent}(a) and Fig.~\ref{Fig:QD}(a) we show $\delta E$ and $\delta D$ as a function of $\tau_{1}$, respectively, for different values of $g/\omega_{c}$. As can be seen by comparing Fig.~\ref{Fig:WEP}(a) with Fig.~\ref{Fig:QD}(a) and ~\ref{Fig:Ent}(a), the total work extracted follows the time dependence of the difference of quantum correlations of the isochoric stages. Notice that while both correlation measures capture the optimal behavior of the engine in terms of the coupling strength $g/\omega$, it is quantum discord which better captures the behavior of the total work extracted, as can be seen for the cases of $g/\omega=0.3$ and $g/\omega=0.9$. This indicates that for this model, quantum correlations other than entanglement have an effect on the performance of the engine. This result can be regarded as generalizing the results obtained for cycles in which the system thermalizes during the heat strokes~\cite{Altintas2014,Hardal2015,Alvarado2017Role}.\par 
In Fig.~\ref{Fig:Ent}(b) and Fig.~\ref{Fig:QD}(b) we plot $\delta E/\tau$ and $\delta D/\tau$, respectively, which are the difference of quantum correlations in the isochoric stages over the total cycle time. As can be seen by comparing Figures ~\ref{Fig:WEP}(b) with ~\ref{Fig:Ent}(b) and ~\ref{Fig:QD}(b), the quantities $\delta E/\tau$ and $\delta D/\tau$ are indicative of the behavior of the power $P$, where again, quantum discord acts as a better indicator. This indicates that the performance of the engine is affected by quantum correlations other than entanglement. 

\begin{figure}[t!]
 	\includegraphics[width=0.9 \linewidth]{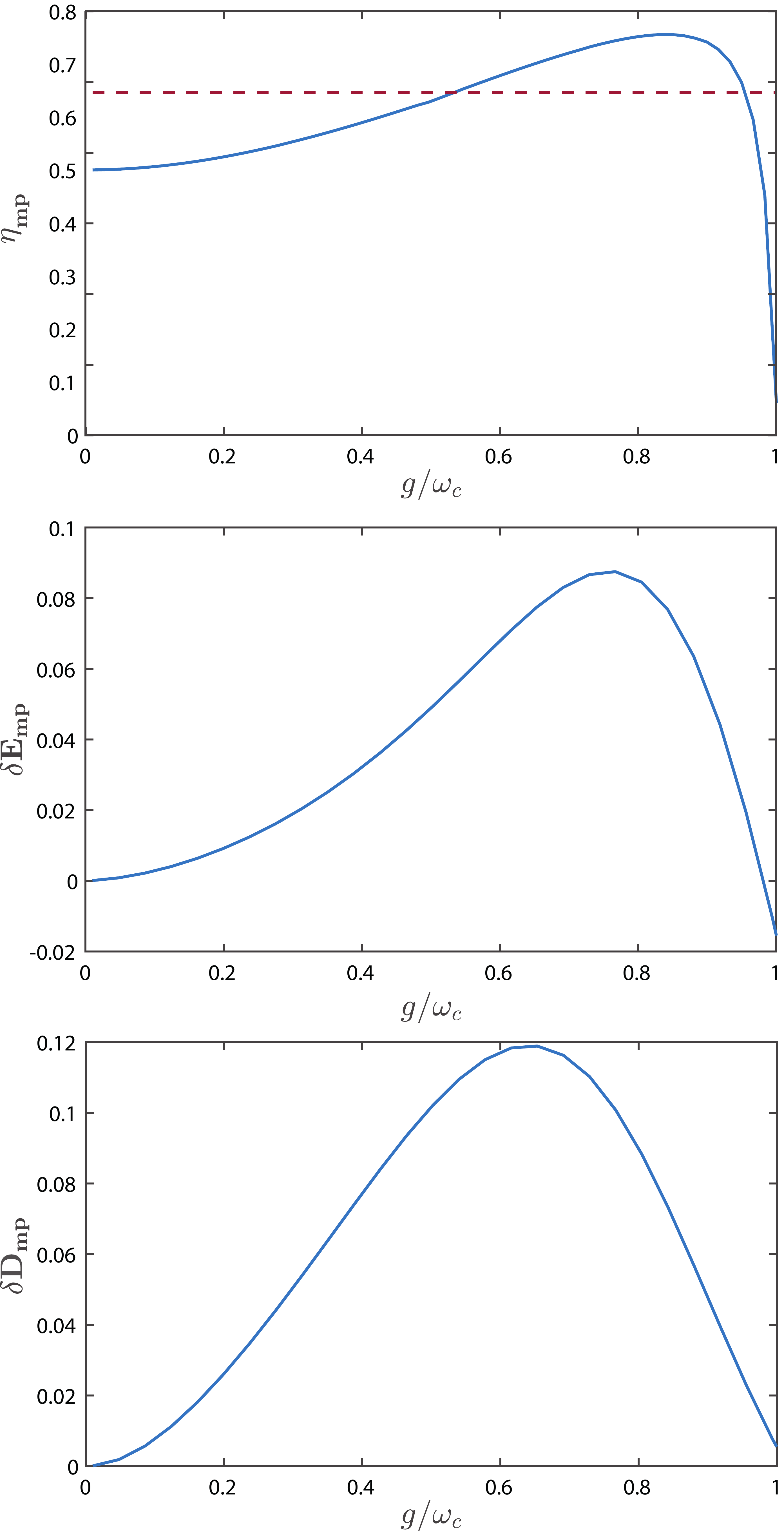}
 	\caption{ (a) Efficiency at maximum power $\eta_{\text{mp}}$ (solid line) and efficiency of Curzon-Ahlborn (dashed line), (b) difference of entanglement $\delta E_{\text{mp}}$ and (c) difference of quantum discord $\delta D_{\text{mp}}$ as a function of the coupling strength $g/\omega_{c}$, at cycle time $\tau_{\text{opt}}$. We have considered $T_{h} = 160$ mK.}
 	\label{Fig:Efficiency_at_MP}
\end{figure}

\section{Efficiency at maximum power}
Finally, we are interested in analyzing the EMP, $\eta_{\text{mp}}$, and its relation with the  change in quantum correlations in the hot isochoric stage. For each value of the coupling parameter $g/\omega_{c}$ we search for the time duration of the cycle, $\tau_{\text{opt}} = 2 \tau_{1}^{\text{opt}} +  \tau_{2}^{\text{opt}}  +  \tau_{4}^{\text{opt}}  $, at which the power is maximum, and calculate the efficiency, $\eta_{\text{mp}}$, for that cycle duration. In $\textrm{Fig.~\ref{Fig:Efficiency_at_MP}(a)} $ we show the EMP (solid line) in terms of the coupling strength $g/ \omega_{c}$, together with the value of the CA efficiency, $\eta_{\text{CA}}$ (dashed line). Remarkably, we notice that for coupling strength values $0.55 \lesssim g/\omega_{c} \lesssim 0.95$, belonging to roughly the USC regime  and near DSC regime, the EMP surpasses the standard bound of CA efficiency, which highlights the capabilities of highly correlated working substances. 

We wish to study if there is a relation between the EMP and quantum correlations, to do so we define the following quantities; we denote the difference of quantum correlations between the initial and final states of the hot isochoric stage by {$\delta \Lambda_{\text{mp}} =  \Lambda(\rho(\tau_{4}^{\text{opt}})) - \Lambda(\rho(\tau_{1}^{\text{opt}}))$}, where $\Lambda(\rho) = \{D(\rho), E(\rho) \}$.  The subindex `mp' indicates that the quantum correlations are calculated at the times $\tau_{i}^{\text{opt}}$ for which the power is maximum. In Fig.~\ref{Fig:Efficiency_at_MP}(b) we show the difference of quantum discord $\delta D_{\text{mp}}$ and Fig.~\ref{Fig:Efficiency_at_MP}(c) the difference of entanglement $\delta E_{\text{mp}}$  as a function of the coupling strength $g/\omega_{c}$. We have plotted these quantities for the range $g/ \omega_{c} \in [0, 1]$ since it is where the total work extracted is positive and hence the efficiency is well defined.  As can be seen by comparing $\textrm{Fig.~\ref{Fig:Efficiency_at_MP}(a)} $, Fig.~\ref{Fig:Efficiency_at_MP}(b) and Fig.~\ref{Fig:Efficiency_at_MP}(c), the difference of entanglement acts as a better indicator of the behavior of the EMP, rather than the difference of quantum discord.

The quantity $\delta E_{\text{mp}}$ (and similarly for $\delta D_{\text{mp}}$) compares the amount of entanglement in the states before and after interacting with the reservoir in the hot isochoric stage. A bigger value of $\delta E_{\text{mp}}$ means the interaction with the reservoir destroyed a greater amount of quantum correlations in the working substance. Since the parameter range of high efficiency roughly coincides with the range where $\delta E_{\text{mp}}$ is greatest, we may interpret that the environment must spend  energy to reduce the quantum correlations in the working substance which can later be harvested by the engine and improve its performance. 

This suggests that while quantum discord may act as an indicator of total work extracted and power, entanglement acts as an indicator of the efficiency of the engine in the optimum regime (maximum energy extracted per unit of time).

\section{Conclusions}
We have studied the finite-time operation of a quantum Otto cycle embedding a working substance composed of a two-level system interacting with a resonator, described by the quantum Rabi model. We obtained the total work extracted, efficiency and power for the limit cycle. We focused our study on the relation between quantum correlations and the total work extracted, as well as the efficiency at maximum power, by considering entanglement of formation and quantum discord. We found that when the hot isochoric stage reduces the quantum correlations in the working substance, it can lead to enhanced positive work extraction. Furthermore, quantum discord, rather than entanglement, acts as a better indicator of the behavior of the total work extracted. On the other hand, we studied the behavior of the engine for a cycle duration which yields the maximum power, and its relation with the quantum correlations in the working subtance. Interestingly, when the working substance is roughly in the range of ultrastrong coupling, it overcomes the Curzon-Ahlborn efficiency. Consequently, it roughly coincides with the cases where the entanglement in the working substance experiences the greatest reduction in the hot isochoric stage. Our results suggest that quantum correlations,~\textit{i.e.}, entanglement and quantum discord, can act as indicators of the performance of a QHE working at finite time, and possibly both should be considered when searching for the optimal configuration. Our study highlight the capabilities of highly correlated quantum systems for enhancing the performance of QHEs for finite time.\par

\section{Acknowledgments}
The authors acknowledge support from ANID Fondecyt, Iniciaci\'on en Investigaci\'on 2020 grant No. 11200032, USM-DGIIE, Universidad de Santiago de Chile, USA2055\_DICYT, Spanish Government PGC2018-095113-B-I00 (MCIU/AEI/FEDER, UE), Basque Government IT986-16, as well as from QMiCS (820505) and OpenSuperQ (820363) of the EU Flagship on Quantum Technologies, EU FET Open Grant Quromorphic (828826), EPIQUS (899368) and Shanghai STCSM (Grant No. 2019SHZDZX01-ZX04).

\bibliographystyle{apsrev}

\end{document}